\documentclass[12pt]{article}
\usepackage{graphics}

\def\cA{{\cal A}}

\def\cC{{\cal C}}

\def\cG{{\cal G}}

\def\cI{{\cal I}}

\def\cK{{\cal K}}

\def\cS{{\cal S}}

\newfont{\goth}{eufm10 scaled \magstep1}

\def\a{\alpha}

\def\b{\beta}

\def\c{\gamma}
\def\d{\delta}
\def\e{\epsilon}
\def\f{\phi}

\def\m{\mu}
\def\n{\nu}

\def\r{\rho}
\def\s{\sigma}
\def\t{\tau}
\def\th{\theta}

\def\beq{\begin{equation}}\def\eeq{\end{equation}}
\def\beqa{\begin{eqnarray}}\def\eeqa{\end{eqnarray}}
\def\barr{\begin{array}}\def\earr{\end{array}}

\def\del{\partial}

\def \ys {{y\kern-.5em / \kern.3em}}

\let\bm=\bibitem

\def\nn{\nonumber}
\def\bd{\begin{document}}
\def\ed{\end{document}}
\def\ba{\begin{array}}
\def\ea{\end{array}}
\def\bea{\begin{eqnarray}}
\def\eea{\end{eqnarray}}
\def\ft#1#2{{\textstyle{{\scriptstyle #1}\over {\scriptstyle #2}}}}
\def\fft#1#2{{#1 \over #2}}
\newcommand{\be}{\begin{equation}}
\newcommand{\ee}{\end{equation}}
\newcommand{\eq}[1]{(\ref{#1})}
\def\eqs#1#2{(\ref{#1}-\ref{#2})}
\def\det{{\rm det\,}}
\def\tr{{\rm tr}}
\newcommand{\ho}[1]{$\, ^{#1}$}
\newcommand{\hoch}[1]{$\, ^{#1}$}
\def\ra{\rightarrow}
\def\uha{{\hat {\underline{\a}} }}
\def\uhc{{\hat {\underline{\c}} }}

\def\bk{{\bf k}} \def\Xh{\hat{X}} \def\ah{\hat{a}} \def\xh{\hat{x}}
\def\ph{\hat{p}} \def\G{{\cal G}} \def\pt{\tilde{p}}
\def\rb{\bar{\rho}} \def\ap{\alpha'}

\newcommand{\sect}[1]{\setcounter{equation}{0}\section{#1}}
\renewcommand{\theequation}{\thesection.\arabic{equation}}
\newcommand{\ex}[1]{{\rm e}^{#1}} \def\ii{{\rm i}}
\newcommand{\hepth}[1]{{\tt hep-th/#1}} 
\newcommand{\hepph}[1]{{\tt hep-ph/#1}} 
\newcommand{\req}[1]{Eq.~(\ref{#1})}
\def\mint{\int\limits}

\begin{document}
\begin{titlepage}

\hfill{NEIP-00-008}

\hfill{hep-th/0003180}

\vspace{20pt}

\begin{center}

{\Large\bf  String Theory and Noncommutative Field Theories at One Loop}
\vspace{30pt}

{\large Adel Bilal, Chong-Sun Chu and Rodolfo Russo}

\vspace{15pt}

{\small \em Institute of Physics, University of Neuch\^atel, CH-2000
Neuch\^atel, Switzerland}

\vskip .2in \sffamily{ Adel.Bilal@iph.unine.ch \\
Chong-Sun.Chu@iph.unine.ch \\ Rodolfo.Russo@iph.unine.ch}
\vspace{60pt}

{\bf Abstract}
\end{center}

By exploiting the boundary state formalism we obtain 
the string correlator between two internal points on 
the one loop open string world-sheet 
in the presence of a constant background
$B$-field. From this derivation it is clear that there is an ambiguity
when one tries to restrict the Green function to the boundary of the
surface. We fix this ambiguity by showing that there is 
a {\em unique} form for the correlator between two points on
the boundary which reproduces the one loop field theory results of
different noncommutative field theories. In particular, we present the
derivation of one loop diagrams for $\phi^3_6$ and $\phi^4_4$ scalar
interactions and for Yang--Mills theory. 
From the $2$-point function we are
able to derive the one loop $\beta$-function for noncommutative gauge
theory. 

\end{titlepage}

\sect{Introduction}

Recently, noncommutative field theory has shown up as an effective
description of string theory in a certain background
\cite{connes,douglashull,CH1,schom,CHL,sw1}. The non-standard
commutation relation among space coordinates takes the form
\be \label{comm} 
[x^\mu, x^\nu] = i \th^{\mu \nu}~~~{\rm with}~~\m,\n\not= 0~, 
\ee 
where $\th^{\mu \nu}$ is an antisymmetric real
constant matrix of dimension length squared. In the dual
language, the algebra of functions is described by the Moyal product
\be 
(f * g) (x) = e^{i \frac{\th^{\mu\nu}}{2} \frac{\del}{\del
\xi^\mu} \frac{\del}{\del \zeta^\nu} } f(x+ \xi) g(x+\zeta)
|_{\xi=\zeta=0}~, 
\ee 
which is associative and  noncommutative.
The noncommutative nature of string theory in the presence of a
non-vanishing $F$-field was emphasized in~\cite{CH1,chudubna}.
There it was shown that when quantizing an open string with the boundary
condition 
\be \label{bc0} 
\del_\s X^\m + \del_\tau X^\n F_\n{}^\m =0,
\ee 
where $F=B-dA$ is the gauge invariant Born-Infeld field
strength, the noncommutativity is an unavoidable feature of
open string dynamics. In fact one finds that
the coordinates $x^\m$ of the open string endpoints 
have to satisfy the new relation  \eq{comm} and also
the commutation relations for the modes of the
string expansion are modified.

Recent interest in noncommutative quantum field theory was
boosted by the paper of Seiberg and Witten~\cite{sw1} where it is
systematically shown how tree-level open string theory, in the
presence of a non-zero $F$-field and of D$p$-branes, leads to a
noncommutative quantum field theory.
One of the main observations in~\cite{sw1} was that, in the presence of
a $F$-field, the open string world-sheet Green's function on the
boundary of the disk is modified~\cite{acny}. In particular, 
Seiberg and Witten showed that in the limit 
\be \label{nclimit} \a' \sim \e,  \quad F \sim 1/\e,  
\ee
\be\label{glimit} \quad g_{ij} \sim \e^2, 
\ee 
the tree level amplitudes {\em just} consist
of a phase factor, which corresponds to the vertices of a
noncommutative field theory. 
The limit $g_{ij} \sim \e^2$ is necessary if one wants to kill
the string propagators to get an
irreducible field theory vertex from an  $M$-point string amplitude. 
In fact, the basic building block in open string theory is
the 3-string vertex: thus in order to get higher point interactions
(e.g. $\phi^{*n}$, $n \geq 4$), one may sew together a couple of
$3$-string vertices and contract some of the propagators in
between. This is accomplished  by \eq{glimit}.  Due to 
some existing
confusions in the literature, we feel that it is worthwhile to
stress that the ``contraction of propagator'' \eq{glimit} should be
imposed only when  necessary and should not always be understood.
Indeed, for our purposes of obtaining more general field theory
amplitudes, it would be wrong to always insist on this limit, 
as this would kill all the propagators,
including the loop propagators. However the other scaling limit
\eq{nclimit} is  necessary to obtain a noncommutative field
theory and we will refer to it as the {\it noncommutative limit}. 
In this paper, we will show that by taking the limit \eq{nclimit} (and
\eq{glimit} only when necessary), 
one can reproduce from string theory different noncommutative field
theories, at the tree and one loop levels. 

The noncommutative scalar 
\cite{chep1,uvir,aref,fischler0}
and gauge theories\cite{g1,g2,susskind,schupp1,anom,chucs}
have been  much studied in their own right.
Intriguing  phenomena occur, in particular there are
important distinctions between planar and nonplanar Feynman diagrams
even in theories of a single scalar. Moreover nonplanar diagrams are
automatically regulated by
an effective UV-cutoff  $\Lambda^2_{\rm eff}\sim 1/(p \th)^2$, where
$p$ is some combination of external momenta. This implies a
non-analyticity in $\th$, and an IR singularity~$\propto 1/(p \th)^2 $ is
generated from integrating out the high momentum modes.  This
UV/IR-mixing has attracted  quite some attention. Within quantum field
theory it appears as a puzzling feature, but if we think in terms of
string theory there is no natural  distinction between UV and IR since
high energy open string  loop excitations are mapped via a modular
transformation to low energy  closed string ones.

It is thus quite natural to suspect that the Seiberg-Witten limit of
string theory with a non-vanishing $F$-field can be extended  beyond
tree-level. In fact, by now the low energy behavior ($\alpha'\to 0$)
of string amplitudes is well understood also at one loop-level and has
provided a reliable and flexible tool for analyzing various aspects of
very different field theories. For instance, string amplitudes or
string--inspired techniques were used to evaluate  one--loop QCD
scattering amplitudes~\cite{beko91,bernrev} 
(see also References therein) and renormalization
constants~\cite{beko88,letter,1loop}; graviton scattering amplitudes
were  computed and their relation to gauge amplitudes
explored~\cite{bedushi}; progress was made towards the extension
of the method to more than one
loop~\cite{kaj,2loop,kajsat,mp,fmr}, 
and to off--shell
amplitudes~\cite{nap}. String techniques also served to stimulate the
development of new techniques in field theory, that preserve some of
the nice features of the  string
formalism~\cite{strass}. Basically,
the flexibility of these techniques has its root in the fact that
string theory has a two-dimensional structure, describing the world-sheet
dynamics beyond the usual space-time structure.  If one can find a
corner of string moduli space which  at low energies yields the field
theory under study, it is possible to use the string description to
perform the calculations and thus exploit all the conformal theory
features one has already studied for other problems. 
Thus, it is not really a
surprise that field theory computations, which are largely independent of
each other, appear to be related if viewed from the string theory
point of view. 

As already discussed above, Seiberg and Witten have
pointed out a regime of string theory which at low energies is
described by a non-commutative theory and thus in the spirit of the
above papers it is natural to exploit string computations to derive
field theory results. Specifically, we will show how the
non-commutative parameter $\theta$ arises in the field theory
limit of one-loop amplitudes of the simple open bosonic string.  We
would like to insist that string theory not only conceptually leads to
the noncommutative quantum field theories,  but it also represents a
simplifying technique for the computation of perturbative
amplitudes. As we will explicitly see, once we have computed 
the objects entering
in the string master formula also in the presence of a
non-vanishing $F$-field, the one-loop diagrams of different 
noncommutative quantum
field theories can be obtained by following the calculations performed in
the commutative case; in particular, we refer to
\cite{1loop,2loop,mp,fmr}. A nice feature discussed in detail in those
references is the existence of a one-to-one correspondence between
Feynman diagrams and different corners of the integration region over the
string parameters. It is worth to stress that this mapping is preserved
in the non-commutative case and is identical to the one found for
$\theta=0$.

In order to incorporate a
non-vanishing $F$-field in one-loop string computations, 
the first non-trivial task is 
to obtain the conformal field theory propagator with $F \ne 0$ on a 
world-sheet with topology of the annulus, rather than simply a disk. This 
will be done in section 2. 
Starting from the boundary state formalism, 
we discuss the possible 
ambiguity for the open string Green function that exists in the
literature and provide an unambiguous computation to fix its form.    
Once this Green function is known, one can apply the techniques
developed in~\cite{1loop,2loop,mp,fmr} in order to extract quantum
field theory Feynman diagrams from string loop amplitudes. 
By means of this formalism, we compute various 
one-loop amplitudes in noncommutative scalar $\phi^3$ and $\phi^4$ field 
theories. In all cases we show that string theory exactly reproduces the 
previously known results obtained from quantizing the noncommutative 
field theory action. We then study the $2$-gluon amplitude 
in noncommutative gauge theories and determine the leading and
subleading singularity in $\th$. By exploiting the fact that in the
field theory limit, string theory gives results in the background
field method~\cite{1loop}, we easily obtain the 
$\b$-function for the noncommutative $U(1)$ gauge theory. 

{\bf Note added:} After completing this work and while the present
paper was typed, two related papers
\cite{nn1,nn2} appeared where also one-loop noncommutative field
theory amplitudes are obtained from one-loop string theory
amplitudes. However, the Green function of \cite{nn1} 
differs from ours since the
above mentioned ambiguity was differently resolved. In section 2 we 
will argue that to obtain the correct gluon two-point function imposes
our choice for resolving the ambiguity.

\sect{One-loop open string Green function in the 
\\presence of an $F$-field}
  
In this section we focus on the one-loop Green function of bosonic
string theory and, in particular, we want to generalize the usual
calculation to the case where a constant field $F=B-dA$ is present. 
In fact, once the explicit form of the Green function is known,
it is possible to write in a compact form a generic string amplitude
with an arbitrary number of legs. 
The situation is thus very different from the
one in field theory, where each diagram represents an independent
calculation and one has always to start from the very first building
blocks, {\em i.e.} the Feynman rules. As we said, this simplification
is possible because string calculations rely on the world-sheet
structure, which is described by a two-dimensional theory, more than
on the space-time structure. In fact, the $h$-loop bosonic string
amplitude among $M$-tachyons with inflowing momenta $p_1,\ldots,p_M$
can be written as  
\beq
A^{(h)}_M (p_1,\ldots,p_M) = C_h {\cal N}^M
\int [dm]^M_h \, \prod_{i<j} \left[\frac{ \exp \left({\cal
G}^{(h)}_{r_i,r_j} (\rho_i,\rho_j)  
\right)}{\sqrt{V'_i(0) \, V'_j(0)} } \right]^{2 \a' p_i\cdot p_j}~~.
\label{hmastac}
\eeq
Here $C_h$ and ${\cal N}$ are the normalization factors depending on the
world-sheet topology and the string vertices respectively; their
explicit form in terms of the dimensionless string coupling constant
will be given later. The other building blocks of \eq{hmastac} 
have a clear geometrical interpretation: ${\cal G}^{(h)}_{r_i,r_j}$
is the correlator of two world--sheet bosons
located at $\rho_i$ on the boundary labeled $r_i$, and at $\rho_j$ on
the boundary $r_j$; $[dm]^M_h$ is the measure of integration 
over the moduli space for an open Riemann surface with $h$ loops and $M$ 
punctures; 
$V'_i(\r)$ are $M$ projective transformations which define local
coordinate systems around each puncture $\r_i$ .
Here we do not give the explicit expressions of these
quantities in general 
(see for instance~\cite{phi}), but we want to stress that
their definition depends only on the geometrical properties of the
string world-sheet and in general on the two-dimensional conformal
theory living on it. From this point of view it is natural that
different computations are much more related to each other than
in the usual field theory approach. 

Here we want to exploit the great flexibility of this technique in
order to derive the one-loop Feynman diagrams of noncommutative field
theories; from the string point of view noncommutativity is easily
implemented: one changes the commutation relations of the open string
modes~\cite{CH1} or, equivalently, the boundary conditions of the
open string coordinates $X^{\mu}(\sigma,\tau)$. At the tree and
one-loop level this modification basically  only shows up in the 
Green function. This means that the definition of the field theory
limit of the string master formula is not modified by the presence of
$F$; in particular, the mapping between the corners of integration over
the moduli and Feynman diagrams can be read from the calculations of
the usual commutative case~\cite{1loop}. Since in our approach 
all the differences between commutative and noncommutative field theory
are resumed in the string Green function, we want to derive here this
key ingredient from first principle.

\subsection{Boundary state formalism} \label{bds}

In \cite{acny}, among other things, 
the tree level  Green function in the presence of a constant
$F$-field was derived by solving the defining differential equation
$\partial_\rho\,\partial_{\bar{\rho}}\,{\cal G}(\rho,\rho') 
=2\pi \alpha'\delta(\rho,\rho')$
with the following boundary condition
\footnote{Notice that there is a factor of
$i$ different from \eq{bc0} due to a Wick rotation on the world-sheet. }
\beq
\label{bc1}
\partial_{\perp}\,{\cal G}_{\mu\nu}(\rho,\rho') - {\rm i}
F_\mu^{\;\sigma}\partial_{\parallel}\, 
{\cal G}_{\sigma\nu}(\rho,\rho')\Big|_{\rho=\bar{\rho}} = 0 ~~,
\eeq
where $\partial_{\parallel}$ ($\partial_{\perp}$) 
is the derivative parallel (normal) to the world-sheet border.
There is however an ambiguity in this approach: one can always add to
a given solution a constant piece ({\em i.e.} independent of 
the punctures $\r,\r'$) with arbitrary
dependence on $F$ and on the annulus width and obtain
another Green function which gives different
results in the $\a'\to 0$ limit. 
As we will see, these terms play a crucial role in
the field theory limit of noncommutative amplitudes, so it is
important to understand the actual form of the Green function
appearing in the string master formula~(\ref{hmastac}). 
In order to clarify this point, we derive the Green function in the 
boundary state formalism using a simple
trick that reduces the actual calculation to the one encountered in the
usual case $F=0$. 

\underline{Tree level}

Let us consider the correlation function of two {\em closed} string
tachyons on a disk,
from which one may extract   the tree-level Green function $\cG^{(0)}$
by  simply  looking at the term with explicit dependence
on both punctures
\beq\label{tree2o}
A_{2}^{(0)} \sim \langle 0| \ex{\ii p_1\cdot X(\rho,\bar{\rho})} 
\ex{\ii p_2\cdot X(\rho',\bar{\rho}')}  |0 \rangle \; 
d^2 \rho\,d^2 \rho'~= 
\ex{p_1 {\cal G}^{(0)}(\rho,\rho') p_2+   
p_1 \cC^{(0)}(\rho) p_1 
+ p_2 \cC^{(0)}(\rho')p_2}
d^2\rho\,d^2\rho'~.  
\eeq
This same amplitude can be calculated in the boundary state formalism.
In this approach one starts from a world-sheet with the topology of
the sphere and thus the string coordinates $X^\m$ depend on two
independent sets of oscillators $\alpha^\m_n$ and
$\tilde{\alpha}^\m_n$. Then one introduces in the amplitudes a
coherent state $| B\rangle$ (see \cite{liccardo} and References therein)
which basically identifies the left and
the right sector of the closed strings with the appropriate boundary
conditions and thus inserts a boundary on the string world-sheet
\beq\label{tree2c}
A_{2}^{(0)} \sim \langle 0,\tilde{0}|\,{\rm T} \left(
\ex{\ii p_1\cdot X(z,\bar{z})} \ex{\ii p_2\cdot X(z',\bar{z}')} 
\right) |B(P) \rangle \; d^2 z\,d^2 z'~,
\eeq
where T is the radial ordering. 
The $P$ in the boundary state $| B(P)\rangle$ is put there to emphasize
that in order to sew to a boundary state to a 
given Riemann surface, it has to contain a closed string propagator
$1/(L_0 + \tilde{L}_0 -2)$ \cite{d1}. 
At tree-level its effect in the amplitudes
is just to shift the positions of the external legs, so it does not
modify the form of the Green functions. 
Notice that in
\eq{tree2o} and \eq{tree2c} we have used two 
different ways to label the world-sheet coordinates. This is because
the two approaches naturally give rise 
to different parameterizations of the
string world-sheet. In \eq{tree2o} the poles of open string exchanges
between the two vertex operators are manifest and the fundamental
region is the upper half of the complex plane. On the other hand,
the boundary state calculation is written in the ``closed string
channel'' and the world-sheet is mapped inside the disk of unit
radius.

The advantage of the boundary state 
formulation is that it is very simple to
introduce a constant $F$-field; in fact, one only needs to slightly
modify the identification brought by the boundary state~\cite{bs,dec1}
(here we use the convention of~\cite{dec1} in the particular case
where there are no Dirichlet directions)
\beq\label{idosc}
\Big[(1\!+\!F)^\m_{\;\n} \a^\n_n + (1\!-\!F)^\m_{\;\n} \tilde{\a}^\n_{-n} 
\Big] | B\rangle_{F} = 0 ~~, \forall\, n > 0~.
\eeq
From this identification it is easy to see that the part of the
tachyon vertex depending on the non-zero modes $\tilde{\a}_n$
satisfies the following relation
\beq\label{idver}
:\ex{\ii {p\over 2}\cdot\tilde{X}'(\bar{z})}: | B\rangle_{F} = 
:\ex{\ii {p_\m\over 2}\left({1\!+\!F \over 1\!-\! F}\right)^\mu_{\;\nu}
X'^{\n}({1\over \bar{z}})}: | B\rangle_{F}~,
\eeq
where $X'$ is the oscillator part of the string coordinate without the
zero modes.  
In fact, since we only consider noncompact and Neumann directions
in $|B\rangle$, both $p$ and $\tilde{p}$ vanish on the boundary; thus  
the zero mode contribution can be calculated separately and modifies
only the $F$-independent part of ${\cal G}$.
Using this identification repeatedly, one can reduce \eq{tree2c}
to the the usual computation of the expectation value of four open
string-like vertex operators and, in general, a $N$-point function of
closed string on a disk is equivalent to a $2N$-point amplitude among
open strings. The only difference is that the former $\tilde X^\m$
part of the vertices is evaluated in the unusual image 
point $1/{\bar z}$
and its Lorentz index is contracted with the external momentum
through a non-trivial matrix depending on $F$. After these
considerations, it is easy to find the expression for the
Green function ${\cal G}$ which depends on two different points $z$
and $z'$
\beq\label{gctree}
{1\over \alpha'}\,{\cal G}^{(0)}_{\mu\nu}(z,z') = 
\delta_{\mu\nu} \ln |z - z'| +
{1\over 2} \left({1\!-\! F \over 1\!+\!F}\right)_{\mu\nu}
\!\!\!\ln (1 - z \bar{z}') +
{1\over 2} \left({1\!+\!F \over 1\!-\! F}\right)_{\mu\nu}
\!\!\!\ln (1 - z' \bar{z})~.
\eeq
Here the antisymmetric nature of the field $F$ has been used to rewrite
the final result in the standard form where the index $\m$ always 
precedes $\n$. It is  clear that ${\cal G}$ is
symmetric under the simultaneous exchange $z\leftrightarrow z'$ and
$\m\leftrightarrow \n$. 

Note that the above result is written in the $z$-coordinates, the
parameterization chosen by the boundary state calculation. 
In order to do the comparison  with the one of~\cite{acny}, it is
necessary to perform a conformal transformation 
$z = - (\rho-i )/(\rho+ i) $
and rewrite the Green
functions in terms of the $\rho$-coordinates. 
A small subtlety in this mapping comes from the fact that the
Green function in \eq{gctree} is not a scalar under conformal
transformations.
The simplest thing to do is to 
transform the scalar combination 
($A_{2}^{(0)} d^2 z\;d^2 z'$) to the $\rho$ parameterization
and read off ${\cal G}^{(0)}(\rho,\rho')$ from there. One finds indeed
the same result ({\em i.e.} Eq.~(2.15) of~\cite{acny}). 

A couple of remarks are in order. 
Notice that the tree-level Green function in the
$z$-coordinates cannot satisfy a boundary condition similar to the one
in \eq{bc1}. This peculiarity has already been stressed 
in~\cite{acny} where it was pointed out that the condition
(\ref{bc1}) may be in contradiction with the equation of motion
because Gauss theorem requires that the sum of all boundary integrals
is equal to $2\pi \a'$. This is indeed the case in
the $z$ coordinate and the boundary condition is modified as
\beq
\label{bc2}
\partial_{\perp}\,{\cal G}_{\mu\nu}(z,z') - {\rm i}
F_\mu^{\;\sigma}\partial_{\parallel}\, 
{\cal G}_{\sigma\nu}(z,z')\Big|_{|z|=1} = \a' ~~.
\eeq
As already anticipated in the amplitudes among
closed string vertices 
on a disk, one also gets a contribution $\cC^{(0)}$
depending on a single point: this comes from the contraction of the
former left and right moving part of the same vertex, since the
oscillators of these two parts are now identified by the presence of
a boundary (\ref{idosc})
\beq
{1\over \alpha'}\,\cC^{(0)}_{\mu\nu}(z) = 
{1\over 4} \left({1\!-\! F \over 1\!+\!F}\right)_{\mu\nu}
\!\!\!\ln (1 - |z|^2) +
{1\over 4} \left({1\!+\!F \over 1\!-\! F}\right)_{\mu\nu}
\!\!\!\ln(1 - |z|^2)~.
\eeq

\underline{One loop} 

The same approach easily generalizes to one-loop, where now two boundary
states must be inserted. Since the noncommutativity we are interested
in at present is related to the global $U(1)$, we choose the 
boundary states to each enforce the same identification on
the oscillators sets of closed strings, 
\beq\label{1-loop2c}
A_{2}^{(1)} \sim {}_F\langle B(P)| \,{\rm T}\left(
\ex{\ii k_1\cdot X(z,\bar{z})} \ex{\ii k_2\cdot X(z',\bar{z}')} 
\right)|B(P)\rangle_{F} \;d^2 z\,d^2 z'~.
\eeq
Again, one can use \eq{idver} in order to eliminate the dependence of
the two vertices on the right moving oscillators, so that the scalar
product $\langle B|B\rangle$ in this sector transforms into a trace
over the remaining left moving $\alpha_n$'s. The final evaluation of the
trace is most easily performed by using coherent states and canonical
forms~\cite{phi}; let us here report and comment on the three basic pieces
of the result. First the measure. 
The two propagators $P$ in \eq{1-loop2c} combine and
give rise to the usual $q^{2L_0}$ factor present in one-loop
amplitudes. It is clear from this point of view that the contribution
to (\ref{1-loop2c}) not appearing in the exponent depends trivially on
$F$ only through the normalization of the boundary state
\cite{dec1}. This part is absorbed in the definition of the measure
(the relation between $[d\m]$ and $[dm]$ is given below)
which thus becomes
\beq\label{meas}
\Big[d\m\Big]_{h=1}^{M=2} = {\rm det}(1+F) 
\Bigg[{d\,q\over q^3}\; d^2 z\; d^2 z' 
\prod_{n=1}^\infty\left({1-q^{2n}}\right)^{2-d} \Bigg]~.
\eeq
Notice here that there is no $(\ln q)^{d/2}$ in $[d\m]$ since we did
not have to perform any Gaussian integral in the closed string
channel. 
Next, by looking at the exponent of the result
\eq{1-loop2c}, one can extract the one-loop Green function
\beqa\label{gc1}
{1\over \alpha'}\,{\cal G}^{(1)}_{\mu\nu}(z,z') & = & 
\delta_{\mu\nu} \left[\ln|z - z'| + 
\ln \prod_{n=1}^\infty  {\left|1-q^{2n}\,{z'\over z}\right| 
\left|1-q^{2n}\,{z\over z'}\right|\over (1-q^{2n})^2}
\right]  \\ \nonumber & + &
{1\over 2} \left({1\!-\! F \over 1\!+\!F}\right)_{\mu\nu}
\left[\ln(1 - z \bar{z}')+
\ln \prod_{n=1}^\infty   {\Big(1-q^{2n}\,{z \bar{z}'}\Big) 
\left(1-{q^{2n}\over z \bar{z}'}\right)
\over (1-q^{2n})^2} \right]  \\ \nonumber & + &
{1\over 2} \left({1\!+\!F \over 1\!-\! F}\right)_{\mu\nu}
\left[\ln(1 - z' \bar{z}) +
\ln\prod_{n=1}^\infty  {\Big(1-q^{2n}\,{z' \bar{z}}\Big)
\left(1-{q^{2n}\over z' \bar{z}}\right)
\over (1-q^{2n})^2} \right]~.
\eeqa
Finally as in the tree-level case, the full amplitude $A_2^{(1)}$ contains
also contractions between the left and the right part of a single vertex
which are encoded in the following $\cC^{(1)}$
\beqa\label{gtc1}
{1\over \alpha'}\, \cC^{(1)}_{\mu\nu}(z) & = & 
{1\over 4} \left({1\!-\! F \over 1\!+\!F}\right)_{\mu\nu}
\left[\ln(1 - |z|^2)+
\ln \prod_{n=1}^\infty {\Big(1-q^{2n}\,{|z|^2}\Big) 
\left(1-{q^{2n}\over |z|^2}\right)
\over (1-q^{2n})^2} \right]  \\ \nonumber & + &
{1\over 4} \left({1\!+\!F \over 1\!-\! F}\right)_{\mu\nu}
\left[\ln(1 - |z|^2) +
\ln \prod_{n=1}^\infty {\Big(1-q^{2n}\,{|z|^2}\Big)
\left(1-{q^{2n}\over |z|^2}\right)
\over (1-q^{2n})^2} \right]~.
\eeqa

We note that \eq{gc1}  
is exactly the Green function obtained in~\cite{acny} (Eq.~(3.6) there), 
which satisfies a certain particular form of boundary condition. 
We remark that
the result \eq{gc1} are written in the ``closed string
channel'' and the natural modular parameter $\ln q = \ii\pi \t_c$
is related to the length of the surface viewed as a cylinder. In this
case a fundamental region for the string world-sheet is the annulus
with inner radius $q$ and outer radius 1. The result of \cite{acny} is
obtained by setting  
$q=a/b$ and rescaling the coordinate $z \to z /b$.

In order to extract from the string amplitude the
contribution of the ``open string channel'', where the world-sheet
degenerates into a circle, one has to perform a modular transformation
on both the Green function and the measure. In particular, at one-loop
level the relation between $z$ and $\rho$-coordinates is~\cite{GSW}
\beq\label{rz1}
z=\ex{2\pi \ii{\ln\rho\over\ln k}}~~,~~~ 
\ln{q} = {2\pi^2 \over \ln k}~.
\eeq
Notice that this identification fixes the cut of the log function in
the  complex $\r$ plane. In fact we want that the segment $(-1,-k )$
of the negative real axis is mapped by \eq{rz1} on the inner border of
the $z$-parameterization; thus we take 
\footnote{Of course, insisting on $    - \pi\leq    \arg \r <\pi$ makes
the logarithm a single valued function at the expense of continuity.}
\be \label{cut}
\ln \rho = \ln|\r| + i \arg \r, \quad        - \pi\leq    \arg \r <\pi
\ee
However, before explicitly performing the modular transformation
(\ref{rz1}) on the various building blocks of the string amplitude, we
want to make two important remarks about the Green function \eq{gc1}. 

First as one can see from \eq{hmastac}, the string
amplitude does not contain simply ${\cal G}$, but its combination
with the derivatives of the local coordinates around each punctures
$V'(0)$~\cite{phi}, and this combination has conformal weight
zero. Usually the $V'(0)$ dependence drops out on-shell, since the
factor coming from the exponent cancels against the one present in the
definition of the the measure \cite{2loop}
\be
[dm]^M_1=[d\m]^M_1 \prod_{i=1}^m
1/V_i'(0) \ .
\ee 
However, in order to exploit the off-shell continuation
of the string results which is possible in the field theory limit
\cite{1loop,2loop}, it is more useful not to perform this
simplification. Thus we use in the amplitudes the measure $[dm]$ instead
of the one of \eq{meas}, and a shifted Green function
\beq\label{Gs}
{\bf G}^{(1)}_{\mu\nu}(z,z') = {\cal G}^{(1)}_{\mu\nu}(z,z') - 
{\a'\over 2} \delta_{\m\n}\left( \ln|V'_z(0)| + \ln|V'_{z'}(0)| \right)
= {\cal G}^{(1)}_{\mu\nu}(z,z')
 - {\a'\over 2}  \delta_{\m\n} \ln|z z'| ~. 
\eeq
Here, as in~\cite{1loop}, we have related the derivative of the
local coordinate to the one-loop Abelian differential $\omega$
since this is the only well defined object on the
annulus having conformal dimension 1. In particular we have 
$V_i'(0) = z_i$. 

The second remark is related to an ambiguity in the determination
of ${\cal G}$ from the boundary state approach. In fact, as we have
already seen, the computation of the amplitude (\ref{1-loop2c}) always
gives a combination of ${\cal G}$ and $\cC$. Thus,
by exploiting momentum conservation, it is possible to shift
terms of a particular form between $\cG$ and $\cC$.
In particular one can extract from the closed string interaction on the
annulus a different definition for ${\cal G}$ and $\cC$
which is still compatible with the final result of \eq{1-loop2c}
\beqa\label{gc2}
G^{(1)}_{\mu\nu}(z,z')&  =& {\bf G}^{(1)}_{\mu\nu}(z,z') 
+M_{\mu\nu}(z,z')
\\ \label{gct2} 
C^{(1)}_{\mu\nu}(z)&  =& 
\cC^{(1)}_{\mu\nu}(z) 
- N_{\mu\nu}(z)~.
\eeqa
The general 
form of this ambiguity is the addition to ${\bf G}_{\mu\nu}$ of a term
$ M_{\mu \nu}(z,z')$ with $M$ satisfying the Laplace
equation, as well as
$M_{\m\n} (z,z') = M_{\n\m}(z',z)$ in order to preserve the
exchange symmetry discussed after \eq{gctree}.
Moreover, in order to ensure that 
a shift of the form \eq{gc2}, \eq{gct2} does not change
the string amplitude  for all mass levels, one also needs the
following properties 
$M_{\m\n}(z,z')  =- M_{\n\m}(z,z')  $, 
$\del_z \del_{z'}M_{\m\n} =0$. Thus $M$ and $N$ are related by
$M_{\mu\nu}(z,z') = N_{\mu\nu}(z)-N_{\mu\nu}(z')$.
This freedom in the definition of the Green function
also appears in the calculation of~\cite{acny} where ${\cal G}$
is derived by solving the Laplace equation on the world-sheet. 
In fact, as in the tree level case, also here it is not possible to
strictly impose on the Green function the same boundary condition
imposed on the string coordinates \eq{bc0}; and hence
there is a certain degree of freedom in the choice of what constraint
is satisfied by $G$. Indeed, the shift in (\ref{gc2}) 
corresponds simply to a redefinition of the boundary condition
of the Green function which leaves unmodified the value of the integral
$\oint \partial_\perp G ds =2 \pi \a'$ that is fixed by 
Gauss' theorem. We
stress that $G$ and $C$ give the same results as those 
obtained with
${\bf G}$ and $\cC$ when they are used
in the contraction of closed string fields $X(z)$ (where thus $z$
must be a point {\em not} located on the boundary of the surface).
However, when one wants to restrict the Green function to the boundary in
order to calculate open string amplitudes, the two Green functions 
may give different results. 

For the case of scalar amplitudes, 
we note that the kind of shift in \eq{gc2} does
not modify the amplitude at all, as it is antisymmetric in $\m$ and
$\n$ and the Green function is contracted only with
external momenta. Therefore the new contributions due to the
additional term sum up to zero using momentum conservation. 
Thus, the two Green functions $G^{(1)}$ and
${\cal G}^{(1)}$ actually give the
same result for tachyon amplitudes. However this is not the case for
gluon amplitudes as the string master formula will involve derivatives
of the punctures, see \eq{gluonmaster} below. Using this 
in the next subsection, 
we will find  that the correct Green function is given by \eq{gc2}. 

\subsection{One-loop open string Green function  with $F\neq 0$}

We begin by evaluating the Green function ${\bf G}^{(1)}_{\mu\nu}(z,z')$
of eq. \eq{Gs}, as obtained from the
boundary state approach, when the arguments take values on any of the two 
boundaries. 
Then we will argue that we need to exploit the above-mentioned 
ambiguity and shift ${\bf G}^{(1)}_{\mu\nu}(z,z')$ 
to $G^{(1)}_{\mu\nu}(z,z')$ 
as in \eq{gc2} in order to correctly 
reproduce the gluon two-point amplitude 
in the noncommutative field theory limit. 

First, we rewrite ${\bf G}^{(1)}_{\mu\nu}(z,z')$ in the following form, 
splitting it into its symmetric and antisymmetric part (in $\mu, \nu$):
\be \label{closedchannel}
{1\over \a'} {\bf G}^{(1)}_{\mu\nu}(z,z') = \delta_{\m\n}\, {\cal I} +
\left({1+F^2\over 1-F^2}\right)_{\m\n}{\cal J} -
\left({F\over 1-F^2}\right)_{\m\n} {\cal K} \; \equiv \cS + \cA ,
\ee
where
\bea
{\cal I} & = &  \ln\left|\sqrt{z/ z'} - \sqrt{z'/ z}
\,\right| +
\ln \prod_{n=1}^\infty {\left|1-q^{2n}\,{z'\over z}\right|
\left|1-q^{2n}\,{z\over z'}\right|\over (1-q^{2n})^2}
\nonumber \\ \label{ijk}
{\cal J} & = &\ln|1 - z \bar{z}'|+ 
\ln \prod_{n=1}^\infty{\Big|1-q^{2n}\,{z \bar{z}'}\Big|
\left|1-{q^{2n}\over z \bar{z}'}\right|
\over (1-q^{2n})^2}\\ \nonumber
{\cal K} & = & \ln\left(
{1 -z\bar{z}'}
\over {1 - {z' \bar{z}}}\right)+ 
\ln \prod_{n=1}^\infty {\Big(1-q^{2n}\,{z \bar{z}'}\Big)
\left(1-{q^{2n}\over z \bar{z}'}\right)
\over {\Big(1-q^{2n}\,{z' \bar{z}}\Big)
\left(1-{q^{2n}\over z' \bar{z}}\right)}}.
\eea
It is easy to see that 
${\bf G}^{(1)}(z,z')$ 
is invariant under $z \to q/z$, $z' \to q/z'$ 
(equivalently $\r \to -k/\r$, $\r' \to -k/\r'$ ) which maps the outer
boundary of the annulus to the inner one and vice versa. 
Note also that ${\bf G}^{(1)}(z,z')$ is single valued on the 
annulus.

As we have already said, the field theory limit of string amplitudes
is more easily performed in the Schoktty representation of the
annulus, since there the open string contributions are
manifest. 
To go to the $\r$ coordinate,
the conformal transformation \eq{rz1} implies the following
transformation
\bea \label{modtransf}
\ln (\sqrt{z/ z'} - \sqrt{z'/ z}\; ) + 
\ln\prod_{n=1}^\infty  
\frac{(1-q^{2n}\,{z'\over z}) (1-q^{2n}\,{z\over z'})}
{(1-q^{2n})^2} +
\ln (\frac{-\pi}{\ln q}) \nn\\
= \frac{ \ln^2 \r / \r'}{2 \ln  k} + \ln (\sqrt{\r/\r'} -
\sqrt{\r'/\r} \;) +
\ln \prod_{n=1}^\infty 
\frac{(1- k^n \frac{\r}{\r'}) (1- k^n \frac{\r'}{\r})}
{(1- k^n)^2}~ .
\eea

Let us discuss the symmetric part of ${\bf G}^{(1)}$ first.
The non-planar open string Green function has one argument on
each boundary, i.e. $|z|=1$ and $|z'|=q$ corresponding to $\r>0$ and 
$\r'<0$, or vice versa. For the planar case, a priori, one has to 
distinguish the case where both arguments are on the outer boundary 
($|z|=|z'|=1$ corresponding to $\r,\r' >0$) from the case where both 
arguments are on the inner boundary ($|z|=|z'|=q$ corresponding to 
$\r,\r' <0$).
It is easy to see that for both planar cases ($\r/\r' >0$)
\be
\cI = I_0  - \ln (\frac{-\pi}{\ln q})~,
\ee
where we have separated the $ln(lnq)$ term, which will eventually
combine with the modular transformation of the measure, from 
the usual Green function $I_0$
\be \label{p2}
I_0(\r,\r') = \frac{ \ln^2 \left(\r / \r'\right)}{2 \ln k } +
\ln \left|\sqrt{\frac{\r}{\r'}} - \sqrt{\frac{\r'}{\r}} \right|
+  \ln \prod_{n=1}^\infty 
\left|\frac{\left(1- k^n {\r\over \r'}\right)
\left(1- k^n {\r'\over \r}\right) }{(1- k^n)^2} \right| .
\ee

For the nonplanar case we have $\r/\r' <0$
and taking into account \eq{cut},
\be
\ln \r/\r' = \ln |\r/\r' | - i \pi,
\ee
so that $|\ex{\frac{\ln^2 \r / \r'}{2 \ln  k}}|
= \ex{\frac{1}{2 \ln  k} (\ln^2 |\r/\r'| - \pi^2) } $
and hence
\be
\cI = - \frac{\pi^2}{2 \ln  k} + I_0  - \ln (\frac{-\pi}{\ln q})
\ee
with $I_0$ given by 
\be \label{np2}
I_0(\r,\r') = \frac{ \ln^2 |\r / \r'|}{2 \ln  k} +
\ln (\sqrt{\left|\frac{\r}{\r'} \right|} 
+ \sqrt{\left|\frac{\r'}{\r} \right|} )
+ \ln\prod_{n=1}^\infty 
\left|\frac{(1+ k^n |\r/\r'|)(1+ k^n |\r'/\r|) }{(1- k^n)^2} \right| .
\ee

Next one has to transform the other piece 
${\cal J}$ to the $\r$ coordinate,
and then restrict the values of $\r$ and $\r'$ 
to the appropriate boundaries.
While this is straightforward, it is simpler to remark, 
that when first restricting 
$z$ and $z'$ to the boundaries one easily sees that
\be
{\cal J} = {\cal I} + 
\left\{
\begin{array}{ll}
0, & \mbox{planar\ (outer \ or \ inner)} \cr
\frac{1}{2} \ln q,  &\mbox{nonplanar} .
\end{array}
\right.
\ee

As a result, we obtain for the symmetric part
\be \label{clch1}
\frac{(1-F^2)}{2} \cS  =   I_0   - \ln (\frac{-\pi}{\ln q}) +
\left\{
\begin{array}{ll}
0, & \mbox{planar\ (outer \ or \ inner)} \cr
\frac{\th^2}{8 \a'^2 \ln  k} ,  &\mbox{nonplanar}
\end{array}
\right.
\ee
where we have identified
\be
\label{thF1}
\th^{\m\n} = 2 \pi \a' F^{\m\n}. 
\ee
As mentioned before, the noncommutative field limit is
defined by $\a' \to 0$ with $\a'F $ fixed and hence $\th$ is a fixed
quantity in field theory.

Now we turn to the antisymmetric part ${\cal A}$.
We will transform this to the $\r$ coordinates and then evaluate it 
on the boundaries. 
Since 
$z\bar{z}'= \exp\left( {2\pi i\over\ln k}\ln{\r\over \bar{\r}'} \right)$,
one has to be careful to correctly take into account the cut of the 
logarithm. For the planar case with both $z,z'$ on the outer boundary 
($\r,\r'>0$) there is no subtlety. With the choice of the cut \eq{cut}, 
the non-planar cases also present no difficulties as we can always take 
$-\pi\le \arg(\r/\bar{\r}') <0$, and the value of the Green function 
on the boundary is just obtained by analytically continuing the Green 
function from the interior of the $\r$-domain. However, to get the planar 
Green function with both $z$ and $z'$ on the inner boundary, we cannot 
analytically continue the resulting expression without encountering the 
cut, since now the argument of $\r/\bar{\r}'$ would have to 
go from 0 to $-2\pi$. 
The simplest and safest way to get the Green function 
for this case is to observe that $G_{\mu\nu}(\r,\r')$ 
must always be 
periodic under $\r\to k\r,\ \r'\to k\r'$, as well as under
$\r\to -k/\r,\ \r'\to -k/\r'$. 
The second transformation exchanges the 
outer and inner boundary, so that the planar Green function on the inner 
boundary equals the one on the outer boundary.
Thus, rather than insisting on analytic continuation which 
physically is irrelevant here, since only the Green functions 
{\it on the boundaries} are relevant, we insist on the physical 
equivalence of the two boundaries. 

As already pointed out, ${\cal A}$ has an ambiguity 
that cannot be fixed in the boundary state computation.
This ambiguity does not affect the scalar amplitudes as
it gives a vanishing contribution when substituted into the string
master formula \eq{hmastac}. However there is an important difference
in the gluon amplitude. Indeed, the gluon master formula \eq{gluonmaster}
contains terms that depend on the 
derivatives with respect to the punctures and there is only one form
of the Green function that can give the correct field theory
result. The ambiguity can be most easily fixed by looking at 
the gluon 2-point function.
We recall from field theory that the planar gluon
2-point function is independent of $\th$. 
It is easy to see that if $|z|=|z'|=1$, 
${\cal K}^{\rm P} = ln(-z/z')$ and using this in the master 
formula would lead to a $\th$-dependence of the planar 2-gluon 
amplitude.  Indeed on the boundary we have
\beq
\cK^{\rm P} = \pm \ln\left({-z\over z'}\right)~~,~~~
\cK^{\rm NP}=0~,
\eeq
where $+$ ($-$) refers to the outer (inner) boundary.
However, the field theory result can be reproduced if 
$\cK$ is shifted as follows
\beq
\cK^{\rm P} = -\ii \pi \epsilon(\r-\r')~~,~~~
\cK^{\rm NP}= \mp {2\pi \ii\over \ln k} \ln|\r\r'|~,
\eeq
which amounts to choose the $N_{\m\n}(z)$ 
introduced in the previous section as
$\pm \a' F/(1-F^2) ln(z)$.
Thus, the final result is 
\be \label{clch2}
\frac{(1-F^2)}{2} \cA =
\left\{
\begin{array}{ll}
- \frac{i \th}{4 \a'}  \e(\r-\r') , & 
\mbox{planar\ (outer \ or \ inner) } \cr
\pm\frac{i  \th}{2 \a' \ln  k} \ln |\r\r'|   ,  &\mbox{nonplanar}
\end{array}
\right.
\ee
where $\e(\r)$ is the step function that is 1 or $-1$ for positive or
negative $\r$.
Note that this open string Green function is to be taken only on the  
boundaries (real $\r,\r'$). Note also that  $G^{(1)}_{\m\n}(\r,\r')$ 
is symmetric under the exchange of the two particles:
\be\label{exchsym}
G^{(1)}_{\m\n}(\r,\r')= G^{(1)}_{\n\m}(\r',\r) \ .
\ee
We see that the only modification in the planar case is 
a step function and it gives rise  
to the usual phase factor of \cite{filk}\footnote{
Note that putting the labels $r$ in
increasing order in clockwise or anticlockwise direction 
is a matter of convention, changing $\th$ to $-\th$ 
and hence in individual Feynman diagrams.
However the total amplitude is always an
even function of $\th$. We will take the convention of clockwise
ordering in this paper.  
}.

Before substituting \eq{clch1} and \eq{clch2} into the string master
formula for the open string amplitude, we note that one has to scale 
them by a factor of  $(1-F^2) /2 $ first. 
The factor of 2 is simply
because a different normalization for the Green function was adopted
for the boundary state formalism and the open string amplitude
(for instance, because of \eq{gc1}, the exponent of \eq{tree2o} is
proportional to $\a' p^2$, while in \eq{hmastac} a factor of $2\a'$ is
present).    
As for  the scaling $(1-F^2)$, it  is
needed when one passes from the closed to the open string amplitudes.
The reason is simple. Let's consider the case of tachyon states whose
vertex operator is $\ex{\ii p\cdot X}$ when $\theta=0$.
As usual one may read the mass of the ground state described
by the above vertex by simply looking at the Virasoro constraint $L_0
-1 =0$.
When the non-commutative parameter is turned on,
the commutation relations for the modes become \cite{CH1,chudubna}, 
\bea
&[a_n^\m, x_0^{\n}]=[a_n^\m, p_0^\n] = [p_0^\m, p_0^\n]=0,
\label{cr1} \\
&[a_m^\m, a_n^\n]=2\a'mM^{-1\m\n}\d_{m+n}
\quad [x_0^\m, p_0^\n]=\ii\,2\a'M^{-1\m\n}, \label{cr2} \\
&[x_0^\m,x_0^\n]= \ii\, 2\pi\a'(M^{-1} F)^{\m\n}, \label{cr3}
\eea
where $M = 1- F^2$ and 
\be
X^\mu =x_0^\mu+(p_0^\mu \tau -  p_0^\nu F_\nu{}^\mu \sigma)+
\sum_{n\neq 0} {\ex{-\ii n\tau} \over n}
\Big(\ii\,a^\mu_n \cos n\sigma -   a_n^\nu F_\nu{}^\mu \sin n\sigma
\Big) ~,
\label{mode1}
\ee
is the mode expansion for the open string coordinates. 
The change in \eq{cr1} and \eq{cr2} is gentle
since it is simply an $F$ dependent rescaling \cite{acny,CZ}, 
\be
\xh_0^\m = x_0^\n (1-F)_\n{}^\m, \quad
\ph_0^\m = p_0^\n (1-F)_\n{}^\m, \quad
\ah_n^\m = a_n^\n (1-F)_\n{}^\m~.
\ee
In terms of these operators, the commutation relations \eq{cr1}, \eq{cr2}
take the standard form, with the $F$-dependence concentrated on
\be
[\xh_0^\m,\xh_0^\n]= 2\pi \ii\, \a'F^{\m\n}. \label{hcr3}
\ee
However since $\xh_0$ does not show up in $L_0$, the
computation of the mass parallels the calculation for $\th=0$ and is
found to be $F$ dependent.
On the other hand, in the field theories we want to reproduce, the
presence of the noncommutative parameter has no effect on the quadratic
part of the Lagrangian, so the mass of the fields do not depend on
$\theta$. In order to reproduce  this feature in the string amplitude, we
rescale quantum numbers like the external momenta and polarizations
by an appropriate factor $p^\m\to (p\cdot (1-F))^\m$.  Notice, as a check,
that this
rescaling exactly absorbs the overall $F$-dependent normalization
of the measure found in the boundary state calculation (\ref{meas}).
Equivalently
one can introduce the hatted variables as in the previous subsection.
With this, the mass of the tachyon takes the usual value $-1/\a'$. 
We remark that this step of rescaling the modes by $ 1-F$ is
equivalent to using a  vertex operator $V= e^{i p\cdot \Xh}$ with 
the  rescaled open string coordinate  $\Xh = X (1-F)$.

As we have mentioned before, we note that there is no
$ln q$ factor in the closed string amplitude just like  the
$\th=0$ case, but there is a power of
$(ln k)^{-d/2} \sim (ln q)^{d/2}$  in the open string amplitude.
There are three sources that these $ln q$ factors can arise when
passing from the closed string to the open string amplitude: from
the modular transformation \eq{modtransf}; from the measure
of integration over the moduli; and from the partition function.
All these factors are independent of $F$ and they combine to give the
desired $ln k$ dependence of the open string  amplitude.

Summarizing,  the open string  Green function with constant
$F$-field in the Schottky representation of the annulus is given by
\be \label{p1}
G_P^{\m\n}(\r,\r') = I_0 \delta^{\m\n} -  \frac{\ii\th^{\m\n}}{4 \a'}
\e(\r-\r') ,
\ee
with
\be \label{p2'}
I_0(\r,\r') = \frac{ \ln^2 \left(\r / \r'\right)}{2 \ln k } +
\ln \left|\sqrt{\frac{\r}{\r'}} - \sqrt{\frac{\r'}{\r}} \right|
+\ln \prod_{n=1}^{\infty}
\left|\frac{\left(1- k^n {\r\over \r'}\right)
\left(1- k^n {\r'\over \r}\right) }{(1- k^n)^2} \right|
\ee in the planar case;
and the nonplanar Green function is
\be \label{np1}
G_{NP}^{\m\n}(\r,\r') = I_0 \delta^{\m\n}
+ \frac{(\th^{2})^{\m\n}}{ 8 \a'^2}\frac{1}{\ln  k}
\pm  \frac{i\th^{\m\n}}{2 \a'} \frac{\ln|\r \r'|}{\ln  k}
\ee
with
\be \label{np2'}
I_0(\r,\r') = \frac{ \ln^2 |\r / \r'|}{2 \ln  k} +
\ln (\sqrt{|\frac{\r}{\r'}|} + \sqrt{|\frac{\r'}{\r}|} )
+  \ln \prod_{n=1}^{\infty}
|\frac{(1+ k^n |\r/\r'|)(1+ k^n |\r'/\r|) }{(1- k^n)^2} | .
\ee
The sign $+(-)$ in \eq{np1} refers to the outer (inner) borders.
Note that $G$ is still symmetric
with respect to the exchange of particles
\be
\label{symm}
G^{\m\n} (\r,\r') = G^{\n\m} (\r',\r)~.
\ee

Exactly these open string Green functions are obtained from the open
string operator formalism  \cite{bbb}.

\sect{String amplitudes in the presence of a constant \\
$F$-field and their field theory limits}

After this detailed discussion of the Green functions and the measure in
the previous section, we now turn to  the actual evaluation of the
one-loop open string amplitudes. In section~\ref{sscal} we start our
analysis by focusing on scalar interactions; 
this is the simplest example since the scalar amplitudes involve
only the ground state of open bosonic string theory (that is the tachyon).
In section~\ref{gsec}, we study the Yang-Mills case.

As a general remark, we want to stress that string amplitudes yield the
correct overall normalization of the various Feynman diagrams without
having to calculate the combinatorial factors typical of field theory
and this agreement holds also in the non-planar case~\cite{fmr}. It is
natural then to expect that also the coefficient of noncommutative
amplitudes are reproduced by the string master formula. We find that
this is indeed the case without having to change the definition of
${\cal C}_h$ and ${\cal N}$. We have~\cite{1loop,2loop,d1,bs,fmr}
\beq\label{norm}
C_h= \frac{1}{(2\pi)^{dh}}\,{g_{\rm op}\,
}^{2h-2}\,\frac{1}{(2\a')^{d/2}}~~,~~~ {\cal N} = \sqrt{2 r}\,g_{\rm
op}\, (2\a')^{d-2\over 4}~, 
\eeq
where $g_{\rm op}$ is the open string coupling constant and $r$ is
related to the normalization chosen for the Chan-Paton factor
${\rm Tr}(\lambda^a\lambda^b) = \frac{1}{r}~\delta^{ab}$. 
Note, in
particular, that the vertex normalization ${\cal N}$ is independent of
the particular string state chosen~\cite{1loop} and will be used to
derive both scalar and ``photon'' interactions. 

\subsection{One-loop amplitudes in scalar theories}\label{sscal}

We are now in the position to derive from the master formula
(\ref{hmastac}) a compact expression that will generate the
noncommutative Feynman diagrams for scalar theories\footnote{Notice
that the overall normalization, in terms of $g_{\rm op}$ is different
from the one of \cite{fmr} because there the convention $r=2$ has been
used. Here we focus on $U(1)$ interaction and thus is more natural to
fix $r=1$.}
\beqa
A^{(1)}_M (p_1,\ldots,p_M) &  = &
{(\sqrt{2}\,g_{\rm op})^M \over (4\,\pi)^{d/2}}
(2 \a')^{(M d - 2 M - 2 d)/4} 
\int\limits_{0}^{1}  \frac{dk}{k} \ex{\a' m^2 \ln{k}} \,
\left( - \frac{\ln k}{2} \right)^{-d/2} 
\nonumber \\ & \times &
\int 
{d \rho_{2}\over \rho_2}~ \cdots \!\! 
\int {d \rho_M \over \rho_M} ~\exp\left[ 
\sum_{i<j} 2\ap p_i^\m p_j^\n G^{(\rm E)}_{\m\n}(\r_i,\r_j)\right]~~.
\label{onemast}
\eeqa
The string projective invariance has been used to choose the fixed points
of the single Schottky generator as $\eta = 0$ 
and $\xi \to \infty$, and to 
fix $\rho_1 = 1$. Also, following~\cite{2loop}, we have to neglect
all $O(k)$ terms in the measure of the integration. In fact they will not
be relevant to the field theory limit where we want to single out the
contributions where only scalar particles (``tachyon'') run in the loop.
Notice also that the original string measure is quadratically divergent
when $k\to 0$, which just signals the presence of a tachyon
instability in the bosonic string. In the field theory 
limit, however, we want to deal with scalars of positive $m^2$, 
so we replace
the tachyon mass $-{1\over \ap} \to m^2$ and $k^{-1} \to {\rm
e}^{\a'm^2 \ln k}$. This may look strange at first sight, but it has
a natural interpretation if one goes back to the old dual
model~\cite{2loop} and is sufficient to reproduce correctly all scalar 
field theory  diagrams.

In the same spirit one does not have to use the full Green function
derived in the previous sections, but can use the effective $G^{(\rm
E)}$ where all positive powers of $k$, corresponding to higher string
modes, have been dropped
\beqa\label{efgs}
&& G^{(\rm E)}_{\m\n\;{\rm P}}\left(\rho_i, \rho_j\right) = 
\delta_{\mu\nu} \left[\ln \left| 
\sqrt{\rho_i\over \rho_j}-\sqrt{\rho_j\over \r_i}\right| + \frac{1}{2
\ln k} \ln^2 \frac{\rho_i}{\rho_j} \right]  
- {\ii\, \theta_{\m\n}\over 4\a'} \e( \r_i -\r_j)~~,
\\ \nonumber
&& G_{\m\n\; {\rm NP}}^{(\rm E)} \left(\rho_i, \rho_j\right) = 
\delta_{\mu\nu} \left[ \ln
\left(\sqrt{|\rho_i|\over |\rho_j|}+\sqrt{|\rho_j|\over |\r_i|}
\right) + \frac{1}{2 \ln k} \ln^2 \frac{|\rho_i|}{|\rho_j|}  \right] 
\pm {\ii\, \theta_{\m\n}\over 2\a'}\; {\ln|\rho_i \rho_j| \over \ln k}
+ {(\theta^2)_{\m\n}\over 8(\a')^2} {1\over \ln k}
~~,
\eeqa
for the planar and non-planar case respectively.

As it is clear from the above expression,
for $\th=0$, the differences
between the planar and nonplanar Green functions are
just some absolute values and signs and it is easy to see that
they lead to the same field theory integrand. This is no
longer true when a non-zero $F$-field is turned on. 
In this case, the non-planar Green function is modified by some
non-trivial terms which depend both on $\th$ and on the moduli
$\rho$ and $k$. As we will see, this modification is enough to
account for  all the differences in noncommutative field theory
between planar and non-planar diagrams.

\underline{$\f^3$ in 6 dimensions}

The first noncommutative field theory we want to reproduce has only
$\phi^3$ interactions:
\beq\label{phi2l}
S_3 = \int {1\over 2} \partial_\m\phi\,\partial^\m\phi -
{1\over 2} m^2\phi^2 + {1\over 3!}\,
g_3~ (\phi * \phi * \phi) ~.
\eeq
By comparing the result for the simplest tree-level amplitude between
string and field theory, one can fix the relation between $g_{\rm op}$
and $g_3$
\be\label{gthree}
g_3=2^{5/2} g_{\rm op} (2\ap)^{d-6\over 4}~.
\ee
Let us now turn to loop amplitudes.
As a warm-up exercise 
we will first compute the one-loop two-point function
in scalar $\f^3$-theory from string theory. This example will show most of
the necessary ingredients. 

By writing \eq{onemast} for the special case $M=2$, one gets
\be\label{adp}
A_2^{\rm P}(p_1,-p_1) = {\ap^{2-d/2}\over (4\pi)^{d/2}} {g_3^2\over 4}
\int_0^1 {d  k\over  k} \,\ex{\a' m^2 \ln{k}} \!
\left( {-\ln  k}\right)^{-d/2}
\int_k^1{d\r_2\over \r_2}~
\ex{\left[-2\ap p_1^\m p_1^\n G_{\m\n}^{\rm P}(1,\r_2)\right]}~,
\ee
since $\r_1=1$ is fixed. The field theory limit ($\a'\to 0$) of
the above expression has to be performed, as usual, by keeping fixed
all quantities that have a meaning in the field theory. 
In particular the logarithmic divergences in the Green function are
related to the dimensionfull Schwinger parameters via a factor of $\a'$.
For instance, one always associates $\ln k$ to the total length of the
loop by taking
\be\label{kt}
\ln  k = -{T\over \ap}~,
\ee
thus $T$ has to be kept finite as $\ap\to 0$. 
In this limit $k$ goes to zero exponentially
which means shrinking the annulus to a one-loop Feynman graph. 
After this replacement the $\a'$ dependence of (\ref{adp}) simplifies and
the whole amplitude is just proportional to a single power of
$\a'$. This means that, in order to have a finite answer for our field
theory limit, it is necessary to introduce one more Schwinger
parameter: since we want to reproduce the irreducible diagram, we
perform the $\a'\to 0$ limit by keeping fixed also $t_2$
\be\label{td}
\ln \r_2=-{t_2\over \ap}
\ee
and get
\be\label{adpp}
A_2^{\rm P}(p_1,-p_1) = {g_3^2\over 4}{1\over (4\pi)^{d/2}} 
\int\limits_0^\infty {dT\over T^{d/2}} {\rm e}^{-m^2 T}
\!\!\int\limits_0^T \!\!dt_2 
\exp{\left[ -p_1^2 t_2 \left( 1-{t_2\over
T}\right)\right]}\; 
\;,
\ee
where we have taken the $\ap\to 0$ limit, since no
overall $\ap$ factor remained.
In this simple case, the $\th$-term does not contribute since $p_1^\m
\th_{\m\n} p_1^\n=0$. \eq{adpp} matches the standard field
theory result written in the Schwinger parameterization, numerical
factor included. Notice that the overall ${1\over
4}$ is half of the usual result determined by  the calculation
of the symmetry factor, because we have restricted ourself to the
planar diagram. 

Now we look at the nonplanar contribution to the 2-point one-loop
open string amplitude. Then $\r_1=1$ and $\r_2\in [-1,- k]$, so 
\eq{td} is replaced by
\be\label{tdnp}
\ln |\r_2|=-{t_2\over \ap} \ .
\ee
Then, using the nonplanar Green function \eq{np1} yields
\beq\label{adnp}
A_2^{\rm NP}(p_1,-p_1) = {g_3^2\over 4}{1\over (4\pi)^{d/2}} 
\int\limits_0^\infty {dT\over T^{d/2}} {\rm e}^{-m^2 T}
\int\limits_0^T dt_2 
\exp{\left[ -p_1^2 t_2 \left( 1-{t_2\over
T}\right)+  p_1^\m p_1^\n {\th^2_{\m\n}\over 4T} \right]} 
\ .
\eeq
Again, 
we are left with the standard Schwinger
proper time integral, but with the additional factor
\be\label{ncco}
{\rm e}^{ {1\over 4T} p_1^\m \th^2_{\m\n} p_1^\n} 
= {\rm e}^{-{1\over 4T} \pt_1^2} \quad {\rm with} \quad
\pt_\m=\th_{\m\n} p^\n~.
\ee
As noted in~\cite{uvir}, for $\pt^2\ne 0$ this serves as an
effective UV cutoff and is at the origin of the UV/IR mixing. 
The amplitude \eq{adnp} exactly coincides with the one obtained from a
direct one-loop 
calculation in noncommutative $\f^3$ field theory.

Let us now consider the irreducible part of the $3$-point amplitude.
In the noncommutative $\f^3$
field theory, there are 2 planar and 6 nonplanar diagrams. 
The 2 planar ones
correspond to the two different cyclic orderings of $p_1, p_2$ and
$p_3$ and are 
each equal to one half of 
the standard commutative diagram times an overall factor
${e}^{-{\ii\over 2} p_1^\m p_2^\n \th_{\m\n}}$ resp. 
${e}^{+{\ii\over 2} p_1^\m p_2^\n \th_{\m\n}}$. 
In string theory we have one
planar diagram with $ k<\r_2,\r_3<\r_1\equiv 1$ which can be splitted 
into
$\r_2<\r_3$ and $\r_2>\r_3$, leading to the 2 different planar
noncommutative field theory diagrams as $\ap\to 0$. We also have a
nonplanar string amplitude where we have 
three choices of which $\r$ is alone on one boundary,
and for each such choice there are two possible orderings 
of the two $\r$'s which are not fixed. Again we will show how the
noncommutative field theory one-loop amplitudes are obtained from the
open string one-loop  amplitude by studying a particular example. The
other diagrams are obtained in a completely
analogous manner. We will start with the nonplanar string loop having
$\r_1=1$ on one boundary and $\r_2,\r_3\in [-1,- k]$ 
on the other boundary.
From the general formula \eq{onemast} we get
\bea\label{atnp}
&& A_3^{\rm NP}(p_1;p_2,p_3) = {g_3^3\over 8} {(2\ap)^{3-d/2}\over 
(4\pi)^{d/2}} 
\int_0^1 {d  k\over  k^2} \left( - \ln  k\right)^{-d/2}
\int_{-1}^{- k}{d\r_2\over \r_2} \int_{\r_2}^{- k} {d\r_3\over \r_3}
\\ \nonumber
&&\times
\exp\left[ 2\ap p_1^\m p_2^\n G^{\rm NP}_{\m\n}(1,\r_2)
+2\ap p_1^\m p_3^\n G^{\rm NP}_{\m\n}(1,\r_3)
+2\ap p_2^\m p_3^\n G^{\rm P}_{\m\n}(\r_2,\r_3)
\right] \ .
\eea
Now let ($-1<\r_2<\r_3<-k$)
\be\label{t3np}
\ln  k = -{T\over \ap}\ , \quad \ln |\r_2|=-{t_2\over \ap} \ , \quad
\ln |\r_3|=-{t_2+t_3\over \ap}
\ee
so that as $\ap\to 0$, $k, \r_2, \r_3$ all go to 
zero exponentially. Again all $\ap$ dependence disappears from the
measure  and after using momentum conservation the resulting amplitude
as $\ap\to 0$ can be written as 
\bea\label{atnpp}
&& A_3^{\rm NP}(p_1;p_2,p_3) 
= {g_3^3\over 8} {{\rm e}^{-{\ii\over 2} p_2\th p_3}\over (4\pi)^{d/2}}  
\int\limits_0^\infty {d T\over T^{d/2}} {\rm e}^{-m^2 T}\!\!
\int\limits_0^T dt_2\!\!\int\limits_0^{T-t_2} \!\!\!dt_3\;
\exp\Big[  p_1\cdot p_2 t_2 (1-{t_2\over T})
\\ \nn
 &&\qquad\qquad
+  p_1\cdot p_3 (t_2+t_3) (1-{t_2+t_3\over T})
+  p_2\cdot p_3 t_3 (1-{t_3\over T}) 
+{1\over 4T} p_1 \th^2 p_1 
- i p_1 \th p_3 {t_3\over T} \Big] \ ,
\eea
where we denote 
$p_1\th p_3 \equiv p_1^\m \th_{\m\n} p_3^\n$ etc. This result
exactly equals the field theory amplitude for the nonplanar Feynman
diagram  with $p_2$ and $p_3$ being
non-planar, as shown in the  Fig~1.
\begin{figure}[ht] 
\begin{center} 
{\scalebox{1}{\includegraphics{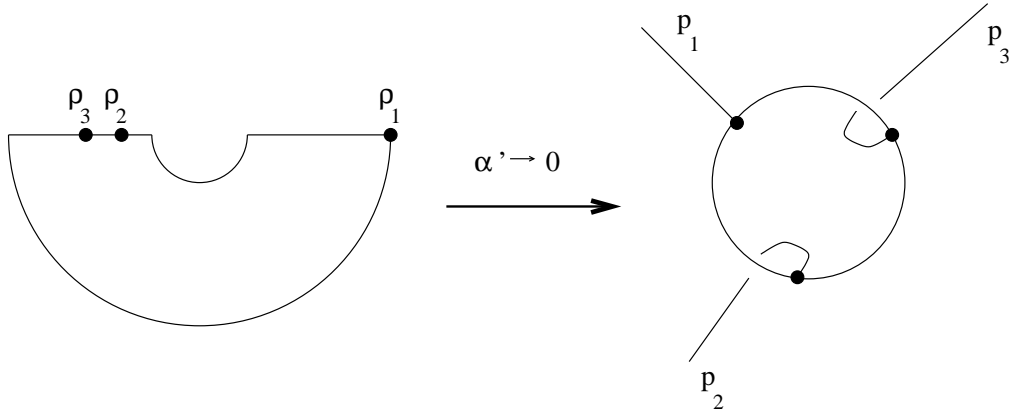}}}
\end{center} \caption{Field theory limit of a nonplanar amplitude.}
\label{btree}
\end{figure}  
The corresponding planar amplitude with the same cyclic ordering,
is easily obtained from \eq{atnpp} by dropping all $\th$-dependent
terms from the exponent inside the integral, but keeping the factor
${e}^{-{\ii\over 2} p_2\th p_3}$ in front of the integral. All the 
other six amplitudes are obtained 
in a completely analogous way and we conclude that
the one-loop string amplitudes correctly reproduce all the 
eight 3-point one-loop diagrams of 
the noncommutative $\f^3$ field theory.

We could now go on and derive higher-point one-loop amplitudes of the 
noncommutative $\f^3$
theory from the string amplitudes but it is pretty clear 
how this will work and that it gives the correct result.

\underline{$\f^4$ in 4 dimensions}

It is now interesting to show that the same string master formula
\eq{onemast} used for $\f^3$ also correctly reproduces the one-loop
amplitudes of noncommutative $\f^4$ theory in $d=4$. 
We follow \cite{mp,fmr,mpp} where $4$-point vertices are obtained by a
different scaling of the string moduli. Again we start by fixing the new
relation between the string coupling constant and the one appearing in
the field theory lagrangian
\beq\label{phi2l4}
L_4 = {1\over 2} \partial_\m\phi\,\partial^\m\phi -
{1\over 2} m^2\phi^2 + {1\over 4!}\,
g_4~ (\phi * \phi * \phi * \phi) ~.
\eeq
Comparing the $4$-point vertex a tree level one can read
\beq\label{gfour}
g_4 = 4!\,(2\a')^{(d-4)/2} g_{\rm op}^2~.
\eeq
where one can immediately check that the critical dimension of $\f^4$ is
correctly reproduced. Since now $g_{\rm op}$ will bring in
\eq{onemast} less powers of $\a'$, in order to have a finite one-loop
amplitude in the field theory limit, one has to consider a limit in
which the distance between  two cubic vertices is vanishing as $\a'\to
0$.

A first simple example is to check how the non-planar $\f^4$ 2-point function
follows from \eq{onemast}. In the field theory diagram we have only one
Schwinger parameter which is related in the usual way to the length
of the string annulus through \eq{kt}. No more rescaling of string moduli
are needed, since the overall factor is now independent of $\a'$ and
thus we shrink to zero the second propagator which would have been
present in the $\f^3$ $2$-point: $-1<<\rho_2 \to 0$, but with $\a'
\ln|\r_2|\to 0$. As expected, this corner of the region of integration
gives the correctly normalized non-planar diagram
\beq\label{1qnpt}
A^{\rm NP}_2(p_1,-p_1) = {g_4 \over 6} {1\over (4\pi)^{d/2}} \int_0^\infty 
{dT \over T^{d/2}}\ex{-m^2 T} \exp{
\left(p_1^\m p_1^\n {\th^2_{\m\n}\over 4T}\right) }~.
\eeq
The planar result is similar to the above equation; a first
difference, of course, is the absence of the $\theta$ dependent
factor; however in this case also the overall coefficient is different
and the factor of $1/6$ is replaced by $1/3$.
As already noticed in the commutative case also, this relative factor
of 1/2, between the planar and non-planar contribution, is reproduced by
string theory \cite{mp,fmr}. In fact in the planar amplitude, on can
consider the region of $\rho_2$ ($\rho_2\to 1$) which gives in the
$\f^3$ case the tadpole diagram. Again it is possible to shrink the
reducible propagator to zero and produce the same $\phi^4$
diagram. Notice that this possibility is allowed only for planar
diagram and thus in this case the final result is doubled.

In the next examples we obtain the four point one-loop diagrams where
$p_1$ and $p_2$ enter the loop at one vertex and $p_3$ and $p_4$ at the
other, with $p_1, p_2, p_3, p_4$ ordered 
clockwise.
In this case one has to scale the moduli $\r_2, \r_3, \r_4$ in
such a way that no
Schwinger proper time is associated to $\r_2/\r_1\equiv \r_2$ or to
$\r_3/\r_4$. Thus, for the planar diagram, we start with the planar
open-string four-point one-loop amplitude $A_4^P$ and do the following
scalings 
\be\label{scalings}
-\ap\ln \r_3=t_3\ ,\quad {\r_4\over \r_3}\ {\rm and}\ \r_2<<1 \ 
{\rm ,~but~finite},
\ee
implying also
\be\label{scalingsbis}
-\ap\ln \r_4=t_3 + {\cal O}(\ap)\ ,\quad -\ap\ln\r_2 \to 0 \ . 
\ee
From \eq{onemast} we then get for the planar 4-point amplitude
\bea\label{aqp}
A_4^{\rm P}(p_1, \ldots p_4)& =& {g_4^2\over 36} {1\over (4\pi)^{d/2}}
\mint_0^\infty {dT\over T^{d/2}} {\rm e}^{-m^2 T} 
\!\!\!\!
\mint_{\ex{-T/\ap}}^1 \!\!{d\r_2\over \r_2} 
\mint_{-\ap\ln \r_2}^T \!\!\!dt_3 \mint_{\ex{-T/\ap}}^{\ex{-t_3/\ap}} 
{d\r_4\over \r_4}
\nn\\
&\times & 
\exp\left[ \sum_{i<j} 2\ap p_i^\m p_j^\n G^P_{\m\n}(\r_i, \r_j)\right] .
\eea
Note again that all factors of $\ap$ have canceled from the measure. 
Since $\widetilde\r_4\equiv \r_4/\r_3$ is finite it is 
more convenient to replace
$$\int_{\ex{-T/\ap}}^{\ex{-t_3/\ap}} {d\r_4\over \r_4}\ldots \to 
\int_{\ex{-(T-t_3)/\ap}}^1 {d\widetilde\r_4\over \widetilde
\r_4} \ldots$$ so that as $\ap\to 0$ we get
\bea\label{aqpp}
&&A_4^{\rm P}(p_1, \ldots p_4)={g_4^2\over 36} {1\over (4\pi)^{d/2}}
\int_0^\infty {dT\over T^{d/2}} {\rm e}^{-m^2 T} 
\int_0^1 {d\r_2\over \r_2} 
\int_0^T dt_3 \int_0^1 {d\widetilde\r_4\over \widetilde \r_4}
\nn\\
&&\quad\quad
\exp\left[ -(p_1+p_2)^2 t_3 (1-{t_3\over T}) 
-{\ii\over 2}( p_1\th p_2 + p_3\th p_4) \right] \ .
\eea
The integrals on $\r_2$ and $\widetilde\r_4$
are divergent in the $\ap\to 0$ limit and are handled much as we did with
divergence in $k$~\cite{fmr}, that is we 
write ${1/\r_2}={\rm e}^{-{1\over \ap} \ap \ln \r_2}
\to {\rm e}^{m^2 \ap \ln \r_2} \to 0$ as $\ap\to 0$ and similarly for 
$1/\tilde{\r_2}$. Hence
the integrals simply give 1 and we finally get
\be\label{aqppp}
A_4^{\rm P}(p_1, \ldots p_4)={g_4^2\over 36} {1\over (4\pi)^{d/2}}
{\rm e}^{-{\ii\over 2} ( p_1\th p_2 + p_3\th p_4) }
\int_0^\infty {dT\over T^{d/2}} {\rm e}^{-m^2 T} 
\int_0^T dt_3 
\ex{\left[ -(p_1+p_2)^2 t_3 (1-{t_3\over T}) \right]}~.
\ee
Just like the two-point amplitude, 
one can also get  contributions from other regions of 
the string moduli in the present case. It is easy to
see that the reducible $\f^3$ diagrams may degenerate giving the
desired $\f^4$ diagram: this can be done in three different ways
according to where are the reducible propagators in the original cubic
diagram. 
One can also reproduce the integrand of \eq{aqppp} from
regions different from the one considered in \eq{scalings}.
For instance, it is possible to exchange the relative order of the
punctures $\r_3$ and 
$\r_4$ and obtain the same result written above with $p_3$ and $p_4$
exchanged. Notice however 
that the two contributions come from the common
boundary of two contiguous regions ($\r_3<\r_4$ and $\r_4<\r_3$), so
that it is natural to weight each of them with a factor of
one half~\cite{fmr}. A similar observation holds also for the punctures
$\r_1$ and $\r_2$ once one remembers that the points $1$ and $k$ of the
Schottky parameterization are identified and mapped to a single point
on the annulus.
Since we have pinched together two pairs of external legs, this brings
a factor of $1/4$ which exactly cancels the ``degeneracy'' coming from
the contributions related to reducible $\f^3$ diagrams and \eq{aqppp}
really represents the final result for the diagram under study. 

Next we look at the same 4-point function but with 
``$p_3$ and $p_4$ nonplanar'',
i.e. putting the vertex on the inner border of the loop rather 
than on the outer one.
Then one has to start with the non-planar string amplitude 
with $\r_1=1$ and
$\r_2$ on one boundary and $\r_3, \r_4$ on the other, i.e. $\r_3,\r_4\in
[-1,- k]$. Inserting absolute values and 
signs at the appropriate places, the
scalings \eq{scalings}, \eq{scalingsbis} remain the same. 
The final result equals
\eq{aqppp} with the phase factor 
$e^{-{\ii\over 2} ( p_1\th p_2 + p_3\th p_4) }$ 
replaced by 
$e^{-{\ii\over 2} ( p_1\th p_2 - p_3\th p_4) }$ and with 
an additional factor ${\rm e}^{{1\over 4T}
(p_3+p_4)\th^2 (p_3+p_4)}$ inserted in the integrand, again in perfect
agreement with the field theory result. Following the above arguments, one
can also easily work out the global coefficient by adding again the
three contribution coming from the regions related to the reducible
$\f^3$ diagrams.

Our last scalar example is this same 4-point diagram but now with only
``$p_3$ being nonplanar''. Going through the same steps as before, 
one is led to
\bea\label{aqpb}
&&A_4^{\rm NP}(p_1,p_2;p_3;p_4)={g_4^2\over 36} {1\over (4\pi)^{d/2}}
{\rm e}^{-{i\over 2} ( p_1\th p_2 - p_3\th p_4) }
\int_0^\infty {dT\over T^{d/2}} {\rm e}^{-m^2 T} 
\int_0^T dt_3 
\nn\\
&&\quad \quad
\exp\left[ -(p_1+p_2)^2 t_3 (1-{t_3\over T}) + {1\over 4T} p_3\th^2 p_3
- i p_3\th p_4 {t_3\over T} \right]~.
\eea
From the string point of view, it is also quite easy to fix the global
normalization by counting how many contributions one can get from the
reducible $\f^3$ diagrams. As is clear in 't~Hooft's
double line notation, two legs can be pulled away and form a
reducible propagator only if they are on the same border. This is also
accounted for 
in string theory where reducible diagrams are related to the
singularity $ln|z_i-z_j|$~\cite{1loop} present only in the planar
case. Thus for this diagram one can get only an additional
contribution beside the one already computed, so that the final result
is twice the one reported in~\eq{aqpb}.

A small remark is due at this point. Contrary to what one expects, 
the results obtained above for scalar interaction do not seem to be
real because of the presence of phase factors. However one has to
remember that the physically meaningful quantity is the whole
amplitude which contains also the contribution coming from the
non-cyclical permutation of the diagrams calculated here. It is easy to
see that in the sum the linear terms in $\theta$ disappear and the
final result is real. However, this observation does not apply to the
last diagram we computed; in fact, in this case, the punctures $\r_3$
and $\r_4$ stay on different boundaries and thus can not be exchanged
without changing the topology of the diagram. This means that the
final result has to be real by itself, 
except for the factor $e^{-\frac{i}{2}p_1 \th p_2}$. 
One can check this by doing 
the change of variables $t_3\to T-t_3$ in \eq{aqpb} whose effect is
exactly to switch the signs of the $p_3\th p_4$ terms.
Of course one or the other form is obtained through a direct
field theory computation.

\subsection{Noncommutative photon} \label{gsec}

Next we study the noncommutative gauge theory with the
action
\be
S=  -\frac{1}{4} \int F_{\mu \nu} * F^{\mu \nu} 
\ee
where $F_{\mu \nu} = \del_\mu A_\nu - \del_\nu A_\mu + i g [A_\mu,
A_\nu]_*$. 
The master formula for reproducing the ``gluon'' amplitude 
can be easily derived from  \eq{hmastac} 
by the usual trick of shifting the 
external momenta by 
$p_i^\m \to p_i^\m + V'_i(0) \,\e_i^\m \partial_{\rho_i}$ 
and then isolating from the result the part linear
in all $\e_i$. We obtain
\bea\label{gluonmaster}
A^{(1)}_M (p_1,\ldots,p_M) &=&
\frac{(\sqrt{2}\,g_{\rm op})^M}{(4\pi)^{d/2}} 
(2 \a')^{(M d - 2 M - 2 d)/4}
\int \limits_{0}^{1}  \frac{dk}{k^2} 
\left( - \frac{\ln k}{2} \right)^{-d/2}\prod_{n=1}^\infty(1- k^n)^{2-d} 
\nn \\ 
& \times & \int
{d \rho_{2}}~ \cdots \!\!
\int {d \rho_M}
~\exp\left[ \sum_{i<j} 2\ap p_i^\m p_j^\n 
G_{\m\n}(\r_i,\r_j)\right]   \\
&  \times & \left[ \exp  \sum_{i\not=j} \left(
\sqrt{2\a'}\, 
\e_i^\mu \,\del_{\r_i}  G_{\m\n}(\r_i,\r_j) p_j^\n 
+\, {1\over 2} \e_i^\mu \, 
\del_{\r_i}\partial_{\r_j} G_{\m\n} (\r_i,\r_j) \e_j^\n 
\right) \right]_{\rm m.l.},  \nn
\eea
where the subscript ``m.l.'' stands for multilinear, meaning
that only terms linear in each polarization
should be kept. As usual, we have used the worldsheet projective
invariance to fix $\delta =0$, $\xi \to \infty$ and $\r_1 =1$. Note that
the measure for the punctures differs from the one of   
scalar amplitude in the expected way. 

The identification of the string coupling with the gluon coupling can
be determined by  comparing the tree level field theory 3-point
function with the tree level string 3-point amplitude and the result is
\be \label{gg}
g = \sqrt{2} g_{op} (2\a')^{(d-4)/4} .
\ee

Before we start  doing  computations using this master formula, we
would like to  comment on the $U(1)$ (non)decoupling in a
(non)commutative  
$U(N)$ Yang-Mills theory from the point of view of string theory. 
In the commutative case, the $U(1)$ factor is free and
decouples. 
As it should be, this can be accounted for in string theory.
For example, there is no $3$-point tree level amplitude in  field
theory. The way string theory produces this fact is quite simple, the 
disc diagram with $3$ vertex operators inserted in one order cancels 
the diagram with the
vertex operators inserted in the reversed order. 
This is a consequence of the fact that vector states of the open 
bosonic
theory carry a quantum number $-1$ under the world-sheet parity
operator $\Omega$. Thus a $3$-point interaction among ``photons'' is
forbidden by the conservation of this quantum number. Note that this 
symmetry argument not only implies that the $3$-point tree level amplitude
is zero, but also forbids  the presence of
internal $3$-``photon'' vertices in more complicated diagrams.
At the loop level, it has been shown in detail~\cite{1loop} that 
string theory again reproduces the one-loop correction to the 
$3$ and $4$-point vertex of Yang-Mills. 
It is instructive to briefly review what is the mechanism that
implements the $U(1)$ decoupling at the one loop.
To understand this, we consider the simple case of  
a $2$-point amplitude where only
one cyclical order of the external legs is possible.  A
vanishing result is now assured by the cancellation among diagrams 
with a different number of legs on the inner and outer border of the
annulus. The reason is quite clear: due to the form of the vector
string state, each vertex brings in the amplitude a derivative of the
Green function with respect of the insertion point of the interaction.
Now since the field theory limit of the 
Green function is an even function 
($G_{\m\n}(\rho)=G_{\m\n}(-\rho)$), 
diagrams with legs on different borders will 
differ  simply by a  change of sign. 
In the $2$-point case this implies that the
planar and the non-planar cancel each other 
to ensure the $U(1)$ decoupling,
even if they are not separately zero. 
In  the noncommutative case, the $U(1)$ factor is not free any more
and does not decouple. This non-decoupling can again be accounted for in
string theory: the various diagrams, which used to cancel each other in
the $\th=0$ case, are now dressed up with different $\th$ dependent 
phases and do not cancel anymore. 
Thus it is clear that the presence of a noncommutative parameter
$\th$ has the same effect as the Chan-Paton trace in the $SU(N)$
case, and it is no surprise to see that a $\th$ dependent
factor here plays the same role as the $SU(N)$ structure constant.

After this comment on the interaction of $U(1)$, 
we are ready to do the computation. We will analyze the gluon
2-point amplitude as an example. 
Restricting \eq{gluonmaster} to the case $M=2$, 
the gluon 2-point amplitude is given by
\bea \label{gl1}
A_2(p_1,p_2) = \frac{(\sqrt{2} g_{op})^2}{(4 \pi)^{d/2}}
(2 \a')^{-1} \int_0^1\frac{d k}{ k^2} 
\left(-\frac{\ln  k}{2}\right)^{-d/2}
\prod_{n=1}^\infty(1- k^n)^{2-d} \int  d \r'
\times \nn\\
\qquad
\; \ex{-2 \a' p_i G^{ij}(\r,\r')p_j } 
\left[
\e_{1i} \frac{\del}{\del \r} \frac{\del}{\del \r'}  G^{ij} \e_{2j}
- 2\a' \left(\e_{1i} \frac{\del}{\del \r}  G^{ij} p^j\right) 
\left(p^i \frac{\del}{\del \r'}  G^{ij} \e_{2j}\right) 
\right]|_{\r=1},
\eea
where $G^{ij}$ is the Green function \eq{p1} or \eq{np1} depending on
whether  we are considering the planar diagram or the nonplanar diagram;
the integration region for $\r'$ is $( k,1)$ in the first case, while
in the latter $ \r' \in(-1,- k)$.
In order to perform the field theory 
limit, as before, one has to introduce the Schwinger parameters 
\be\label{resg}
 k= \ex{-T/\a'}, \quad \r'= - \ex{-t/\a'} 
\ee
with $t,T$ fixed.
The modification in the planar case is simple, the additional term in the
Green function  is a step function  \eq{p1}. In general one should
drop the contribution coming from the derivatives of  the step
function, as it corresponds to put two vertex operators at the same
point, which is not allowed in string interaction. 
Therefore the only modification to the planar $M$ gluon amplitude 
is the usual phase factor.

The nonplanar case is more interesting. Using \eq{np1}
and \eq{np2}, we arrive  at
\be
A_2(p,-p) = \frac{(\sqrt{2} g_{op})^2}{(4 \pi)^{d/2}}
(2 \a')^{-1} 
 \int_0^1 [d k] \int_{-1}^{- k} d\r'
\ex{-2\a' p^2 I_0 - \frac{(\th p)^2}{4T}} \; J ~,
\ee
where $I_0$ is given by \eq{np2},
$[d k]$ is the measure
\be \label{measure}
[d k] = \frac{d k}{ k^2} \left(-\frac{\ln  k}{2}\right)^{-d/2}
\prod_{n=1}^\infty(1- k^n)^{2-d} ~,
\ee
and $J=J_0+J_1+J_2$ are 
the polarization dependent pieces in \eq{gl1}
$$
J_0 = \e_1\cdot \e_2 \frac{\del^2 I_0}{\del\r \del\r'}  
- 2\a' (\e_1\cdot p) (\e_2\cdot p)
\frac{\del I_0}{\del\r } \frac{\del I_0}{\del\r'} ~,
$$
\be
J_1 = \frac{i}{\ln  k} \left[
\frac{\del I_0}{\del\r'}  (\e_1\th p) (\e_2\cdot p) + 
\frac{1}{\r'}\frac{\del I_0}{\del\r}    (\e_2\th p) (\e_1\cdot p)
\right]~,
\ee
$$
J_2 = \frac{-1}{2 \a' (\ln  k)^2} \frac{1}{\r'} (\e_1 \th p) (\e_2 \th
p)~.
$$
where the subscript of $J$ denotes the powers of $\th$.

The 1-loop amplitude, as computed in \eq{gl1}, contains the propagation
of all the string states within the loop. This is partially reflected
in the fact that the measure contains all powers of $ k$. 
As explained in~\cite{1loop}, the contribution from the gluon 
(which is at level 1 of the open string spectrum) can be identified
with the contribution coming from the 
pieces linear in $ k$ in the integrand \eq{gl1}. 
It is straightforward to do the  expansion in $k$  in order to isolate
the relevant factors
\be
\int_0^1 [d k] =  2 \,(2\a')^{ d/2-1 } \mint_0^\infty
\frac{dT}{T^{d/2}} \; \left[ \frac{1}{ k} + (d-2) + \cdots \right] ~,
\ee
\be
\frac{\del I_0}{\del\r'}  = \frac{1}{\r'} 
\left( \frac{t}{T} - \frac{1}{2} - \r' + 
\frac{ k}{\r'}-  k \r' + O( k^2) \right)
= - \frac{1}{\r'} \frac{\del I_0}{\del\r} ~.
\ee
If we now pass from the string moduli to the Schwinger parameters by
means of \eq{resg}, we get 
\be
A_2(p,-p) = \frac{2}{\a'}  \frac{g^2}{(4 \pi)^{d/2}}
\mint_0^\infty
\frac{dT}{T^{d/2}} \; \left[ \frac{1}{ k} + (d-2) + \cdots \right] 
\mint_0^T d t
\;\; \ex{-(2\a') p^2 I_0 - \frac{(\th p)^2}{4T}} \; (\r' J) ~.
\ee
Let us focus now on the contribution which is linear in $\th$
\be
J_1 \sim  
\frac{-\ii}{\ln  k} \frac{1}{\r'} \left( 
\frac{t}{T} - \frac{1}{2} - \r' + \frac{ k}{\r'}-  k \r' +\cdots
\right)~.
\ee
and since $- \r' + \frac{ k}{\r'}-  k \r' \rightarrow 0 $, so 
the contribution of $J_1$ in the two point function  is 
\bea
A_2(J_1) & =& -2\,\ii\, (d-2) \frac{g^2}{(4 \pi)^{d/2}} 
\Big[(\e_1\th p) (\e_2\cdot p) - (\e_2\th p) (\e_1\cdot p)\Big]
\nonumber \\ &\times& 
\mint_0^\infty \frac{dT}{T^{d/2+1}} 
\mint_0^T\! dt\; 
\left(\frac{t}{T}-\frac{1}{2}\right) \;
\ex{-p^2 (t -\frac{t^2}{T}) - \frac{(\th p)^2}{4T}}~.
\eea
It is easy to see that the integral of $t$ 
is is a total derivative and integrates to zero. 
Therefore the linear term $J_1$ is vanishing. 
$$
A_2(J_1) = 0. 
$$

As for $J_0$, one can do an integration by part on $\r'$ and it is
easy to see that one reproduces the usual tensor structure for the 2-point
amplitude. Thus the explicit form of the $J_0$ contibution is
\bea
A_2(J_0) & = & -4\,\frac{g^2}{(4 \pi)^{d/2}} 
\Big[(\e_1 \e_2)p^2 - (\e_1\cdot p) (\e_2 \cdot p)\Big]
\\ \nonumber &\times &
\mint_0^\infty \frac{dT}{T^{d/2}}
\mint_0^T dt
\left[(\frac{t}{T}-\frac{1}{2})^2(d-2) -2 \right]
\ex{-p^2 (t -\frac{t^2}{T}) - \frac{(\th p)^2}{4T}}~.
\eea
For $d=4$, the integral is 
$$
\frac{1}{2}
\mint_0^\infty \frac{dT}{T} \ex{-\frac{\pt^2}{4T}} \mint_0^1 dx
\ex{-T p^2 x(1-x)} [(1-2x)^2 -4] =
\mint_0^1 dx [(1-2x)^2 -4]  K_0(\sqrt{x(1-x)} |p||\pt| )~.
$$
Now the leading  divergence for the Bessel function is
\be
K_0(x) \sim \ln\frac{2}{x} , \quad x\rightarrow 0~.
\ee
So we obtain
\be
A_2(J_0) = \frac{g^2}{(4 \pi)^{d/2}}
\left(\frac{22}{3} \ln( |p|^2 |\pt|^2)  +\cdots \right)
\; [(\e_1 \e_2)p^2 - (\e_1\cdot p) (\e_2 \cdot p)]~.
\ee
Notice that there is a $ln \th$-type singularity.

Finally we look at $J_2$, which is more interesting. It is
\be
J_2 = -\frac{\a'}{2 T^2} \frac{1}{\r'}  (\e_1 \th p) (\e_2 \th p)~,
\ee
and hence it is contribution in $A_2$ is
\be
A_2(J_2) = -(d-2) \frac{g^2}{(4 \pi)^{d/2}}
\int_0^\infty \frac{dT}{T^{d/2+2}} 
\int_0^T dt \ex{-p^2 (t -\frac{t^2}{T}) - \frac{(\th p)^2}{4T}} 
(\e_1 \th p) (\e_2 \th p)~.
\ee
For $d=4$, the integral is equal to 
\bea
\mint_0^\infty \frac{dT}{T^{3}}  \ex{- \frac{(\th p)^2}{4T}} 
\mint_0^1 dx \ex{-T p^2 x(1-x)} 
&=& \frac{16}{\pt^4}  \cdot
\frac{\del^2}{\del a^2} 
( \mint_0^\infty \frac{dT}{T}  \ex{- \frac{a(\pt)^2}{4T}} 
\mint_0^1 dx \ex{-T p^2 x(1-x)} )|_{a=1} \nn\\
&=& \frac{32}{\pt^4}  \cdot
\frac{\del^2}{\del a^2} \mint_0^1 dx
K_0(\sqrt{x(1-x)}|p||\pt|\sqrt{a})|_{a=1}~.
\eea
By expanding the above formula for small momenta, we can extract the
leading IR behavior of $J_2$
\be
A_2(J_2) =  \frac{2 g^2}{\pi^2} 
\frac{ (\e_1 \pt)( \e_2 \pt)}{\pt^4} + \cdots .
\ee
This form of $1/\th^2$ singularity in the two gluon amplitude 
was first obtained in~\cite{g2} and is another example of 
UV/IR mixing in noncommutative field theory which was stressed in
\cite{uvir}. Also in the 4 dimensional scalar $\phi^4$ theory 
the leading singularity is of the form $1/(\th p)^2$ as one can easily see
from the result of the nonplanar $2$-point function \eq{1qnpt}.  
Since the commutative gauge theory is less divergent in the UV than
the commutative scalar theory, one could have expected that the leading 
singularity in the noncommutative gauge theory is of the form 
$ln (\th p)^2$. However, as was
shown in \cite{g2,susskind}, 
this expectation is not correct and one
still gets the  $1/(\th p)^2$ behavior.   
As pointed out in \cite{susskind}, the appearance of the $ 1/(\th
p)^2$  piece does not spoil gauge invariance. In fact it 
does not correspond to a mass shift because of its unusual tensor
structure. 
Here we have re-derived this result using string theory. In addition, we
have been able to determine also the subleading  singularity from string theory
which is of the form  $\ln (\th p)^2$ from string theory.

Summarizing, we have 
\be \label{g2result}
A_{\rm NP}  = \frac{g^2}{16 \pi^2}
\left\{ 32 \frac{ (\e_1 \pt)( \e_2 \pt)}{\pt^4} + 
\frac{22}{3} \ln (|p|^2 |\pt|^2)  
[(\e_1 \e_2)p^2 - (\e_1\cdot p) (\e_2 \cdot p)] 
\right\}
+ \ldots
\ee
where dots stay for terms which are finite when $\th\to 0$.
It is interesting to note that the whole result comes from a single
corner of the moduli space. This is because we performed an
integration by parts which has the effect of including the tadpole
contribution in the region here considered.
We note that the vacuum polarization of the photon  was also obtained by 
Hayakawa \cite{g2} in the Feynman gauge. 
The coefficient of the $ln \th$ term is clearly related to the
wave-function renormalization constant of the commutative case and
thus is gauge dependent. The result found in \cite{g2}
is thus different from ours, because his calculation was performed in 
the Feynman gauge. On the contrary, as explained in~\cite{1loop} for
the commutative case, string theory amplitudes naturally give
rise to field theory results in the background field method~\cite{abbott}. 
Recall that a distinct property of Feynman diagrams
in the background field gauge is to leave the gauge invariance of the
external particles unbroken.  
For the noncommutative case, the situation is the same and 
we conclude that string theory again gives field 
theory results computed in the background field method.
One of the advantages of our result is
that the $\b$ function can be determined 
directly from the two point function without the need 
of computing the vertex corrections. By exploiting this and the fact that
the planar diagram contribution is exactly the same (apart from a
trivial overall group theory factor of $N$) as those in a
usual commutative $SU(N)$ gauge theory, we easily obtain  
the 1-loop $\b$ function
\be
\b = - \frac{g^2}{16 \pi^2} \frac{22}{3}~.
\ee
The $\b$-function for the noncommutative QED has also been computed in
\cite{g1} \cite{g2} 
in the Feynman gauge, by combining vertex corrections and wave
function renormalization. 
The Yang-Mills $\beta$-function can be elegantly derived also from the
string generating functional by following the approach 
\footnote{
We would like to thank A. Tseytlin for pointing out this possibility to 
us.}
of Mastaev and
Tseytlin~\cite{mt}. Again, the noncommutative paramenter $\theta$ can be
easily implemented in the string calculation by switching on a constant
magnetic field. It is easy to see that the derivation of~\cite{mt} is not
substantially modified and one finds the distinctive factor $26-D$: this
is, from this point of view, the source of the Yang-Mills 
$\beta$-function $\sim 22/3$.

\sect{Discussion}

In this paper, we used  the boundary state approach 
to compute
the  one-loop open string  Green function by calculating the 
string amplitude with two boundaries inserted. 
We note that generally
there is  an ambiguity in the resulting Green function 
that cannot be fixed from the
boundary state approach. We pointed out that 
the scalar amplitudes do not care about this
ambiguity. However, this ambiguity can be fixed by simply 
comparing with the planar gluon 2-point function and one obtains the correct
open string Green function which can then be used in the string master
formula in a uniform manner for all mass levels.
A more satisfactory way \cite{work} 
to obtain the desired Green function would be
to calculate directly the open string  
scattering amplitudes using the vertex operators constructed from the 
open string modes \eq{cr1}-\eq{cr3} 
and extract from there the Green function. 

An interesting behavior of noncommutative quantum field theory
is that although classically the noncommutative
description is a smooth deformation of the commutative description,
quantum mechanically they are different and generally 
one cannot turn off $\th$ to smoothly reduce back to the commutative
description. The UV/IR mixing \cite{uvir} and 
the induced Chern-Simons action in
odd dimensions \cite{chucs} are examples. 
The fact that a commutative field theory differ intrinsically 
from a noncommutative field theory is clear
from the present approach. It is important 
to recall that in deriving the noncommutative 
field theory limit from string theory, one has to take the {\it double
scaling  limit} \eq{nclimit} with $\a' \to 0$, $F \to \infty$ and 
$\a' F$ fixed in order to have a nonzero $\th$. The commutative case 
$\th=0$, however,  is obtained from a different limit with $F$ fixed
instead and it is thus clear that the commutative and the
noncommutative field theories thus obtained differ in character.

We now discuss some possible further directions of work.
In this paper, we have shown that by taking the limit \eq{nclimit},
\eq{glimit} appropriately, one can obtain field theory results
from a single string master formula in a uniform manner. We presented
explicit calculations for both
the case of noncommutative scalar theories and noncommutative gauge
theory with gauge group $U(1)$. However, it is straightforward to add
Chan-Paton factors and obtain results for noncommutative field theory
with gauge group $U(N)$. One difference is that in the $U(N)$ case,
the action is no longer even in $\th$ and one will get field theory
result with odd power dependence in $\th$ also. This can definitely 
be accounted for by the Green function we obtained. 
Another possible generalization is to consider superstring amplitudes 
in the presence of $F$-field and to extract the field theory limit. 
The field theory is of course a noncommutative supersymmetric one. It
may be interesting to understand some of the field theory issues (like
non-renormalization theorem and duality for noncommutative field
theory) from the string point of view. Also, 
it is possible to generalize all
this to higher loops \cite{work} and we hope to report on some of 
these topics in the future.

\vspace{.5cm}

\bigskip
\vspace{.5cm}

\noindent{\large \bf Acknowledgments}

R. R. wants to thank P. Di Vecchia, M. Frau, A. Lerda and R. Marotta
for useful discussions and A. Frizzo and L. Magnea for e-mail
exchange.
We are also grateful to H. Dorn for helpful correspondence after the
submission of the first version of the present paper.
This work was partially supported by the Swiss National Science
Foundation, by the European Union under TMR contract
ERBFMRX-CT96-0045 and by the Swiss Office for Education and
Science. 


\ed